\documentclass{aastex61}

\usepackage{amsmath,amstext}
\usepackage[T1]{fontenc}
\usepackage{apjfonts} 
\usepackage{float}
\usepackage{placeins}

\usepackage[figure,figure*]{hypcap}

\usepackage{color, xcolor}

\shorttitle{Fe K$\alpha$ lines from super-Eddington accretion disks}
\shortauthors{Thomsen et al.}

\begin{document}

\title{X-ray fluorescence from super-Eddington accreting black holes}

\correspondingauthor{Lars Lund Thomsen.}
\email{gfh112@alumni.ku.dk}
\author{Lars Lund Thomsen}
\affiliation{Department of Physics, University of Hong Kong, Pokfulam Road, Hong Kong}
\affiliation{DARK, Niels Bohr Institute, University of Copenhagen, Lyngbyvej 2, 4. floor, 2100 Copenhagen {\O}, Denmark}

\author{Jane Lixin Dai}
\affiliation{Department of Physics, University of Hong Kong, Pokfulam Road, Hong Kong}
\affiliation{DARK, Niels Bohr Institute, University of Copenhagen, Lyngbyvej 2, 4. floor, 2100 Copenhagen {\O}, Denmark}

\author{Enrico Ramirez-Ruiz}
\affiliation{Department of Astronomy and Astrophysics, University of California Santa Cruz, 1156 High Street, Santa Cruz, CA 95060, USA}
\affiliation{DARK, Niels Bohr Institute, University of Copenhagen, Lyngbyvej 2, 4. floor, 2100 Copenhagen {\O}, Denmark}

\author{Erin Kara}
\affiliation{MIT Kavli Institute for Astrophysics and Space Research
Massachusetts Institute of Technology
77 Massachusetts Avenue, 37-241
Cambridge, MA 02139}

\author{Chris Reynolds}
\affiliation{Institute of Astronomy, University of Cambridge, Cambridge, CB3 0HA, UK}
\begin{abstract}
X-ray reverberation has proven to be a powerful tool capable of probing the innermost region of accretion disks around compact objects. Current theoretical effort generally assumes that the disk is geometrically thin, optically thick and orbiting with Keplerian speed. Thus, these models cannot be applied to systems where super-Eddington accretion happens because the thin disk approximation fails in this accretion regime. Furthermore, state-of-the-art numerical simulations show that optically thick winds are launched from the super-Eddington accretion disks, and thereby changing the reflection geometry significantly from the thin disk picture. 
We carry out theoretical investigations on this topic by focusing on the Fe K$\alpha$ fluorescent lines produced from super-Eddington disks, and show that their line profiles are shaped by the funnel geometry and wind acceleration. We also systematically compare the Fe line profiles from super-Eddington thick disks to those from thin disks, and find that the former are substantially more blueshifted and symmetric in shape. These results are consistent with the observed Fe K$\alpha$ line from the jetted tidal disruption event, \emph{Swift} J1644, in which a transient super-Eddington accretion disk was formed out of stellar debris. Therefore, careful analysis of the Fe K$\alpha$ line profile can be used to identify systems undergoing super-Eddington accretion.  
\end{abstract}


\section{Introduction} \label{sec:intro}
Some of the most luminous astrophysical sources, such as the active galactic nuclei (AGNs), X-ray binaries (XRBs) and long gamma-ray bursts (GRBs), are all powered by the accretion of gas onto black holes (BHs). Viscous or magnetic process in the gaseous disk  transports angular momentum outwards and heats up the disk \citep{Shakura73,Balbus1991}. In this process, part of the gas mass-energy is effectively converted into energy and released in the form of radiation or large-scale outflows in the wind and jet regions.
Therefore, it is of the uttermost importance to study the detailed structure of accretion disks. While a Keplerian rotating thin disk with no outflows \citep{Shakura73,Novikov1973} is widely applied to explain many accreting systems, this disk model will break down when the accretion rate is very low or very high. For the latter case, it is because there exists a theoretical upper threshold of radiation level, which corresponds to the equilibrium when the radiation pressure force on a gas element balances the gravitational force it receives from the BH. This maximum luminosity, called the Eddington luminosity, $L_{\rm Edd}$, and the corresponding Eddington accretion rate, $\dot{M}_{\rm Edd}$, are given by the following equations
\begin{equation}
    L_{\rm Edd} \approx 1.26 \times 10^{38} \Bigg( \frac{M}{M_{\odot}}\Bigg) \ {\rm erg \ s^{-1}}  \ \ \ {\rm and} \ \ \   \dot{M}_{\rm Edd}= \frac{L_{\rm Edd}} {\eta \ c^{2}},
\end{equation}
where $M/M_\odot$ is the mass of the central object in units of solar masses $M_\odot$, $c$ is the speed of light and $\eta$ is called the radiative efficiency. When the accretion level exceeds this limit, the radiation pressure will become large enough to change the disk structure; making it geometrically and optically thick \citep{Begelman1978, Abramowicz1980}. Recently, great progress has also been achieved through numerical studies with the aid of general-relativistic radiation magnetohydrodynamic (GRRMHD) codes \citep{Ohsuga2009,Jiang2014,McKinney2014, Sadowski2014,Dai18}. These simulations of super-Eddington accretion flows have shown that the Eddington luminosity limit can be broken and large-scale optically-thick winds are launched from the disk. These findings have significant implications on how much of the supplied gas can eventually be accreted onto the BH, how much energy is carried away by outflows and how much radiation can escape. Fast winds, consistent with simulation results, have also been observed from ultra-luminous X-ray sources (ULXs) \citep[e.g.,][]{Pinto2016} and tidal disruption events \citep[e.g.,][]{Kara16,Kara2018, Alexander2017}, both likely undergoing super-Eddington accretion.

Tremendous effort has been put into constraining the structures of accretion disks from observational features. In particular, X-ray reverberation, first proposed by \citet{Fabian89}, has proved to be very useful in giving direct measurement of the inner disk geometry by combining both the spectral and temporal information from this region. As presented in the review by \citet{Reynolds14}, the classical approach to X-ray reverberation is to place a hot and compact corona above the cold accretion disk from where it emits non-thermally distributed X-rays. The disk is illuminated by the coronal emission, which then gives rise to an X-ray reflection spectrum. One notable feature of the reflection spectrum is the Fe K$\alpha$ fluorescent lines that are produced from the irradiated and ionized disk \citep{Matt93}. 
General relativistic (GR) and Doppler effects between the rotating disk and the observer jointly make the fluorescent line profiles broadened and skewed. 
This theoretical model has been successfully applied to constrain the spin of BHs in systems where the disk has a razor thin structure and is rotating with relativistic Keplerian speed \citep[e.g.,][]{Fabian89, Reynolds97,Reynold99, Reynolds14}.
The line profile has a dependence on the BH spin because the innermost edge of a thin disk is given by the innermost stable circular orbit (ISCO) of the BH, which is smaller for a BH with larger prograde spin.
X-ray reverberation has also been used to study the nature of coronal formation, for which a full understanding has not yet been achieved \citep{Blandford2017,Yuan2019}. Therefore, various coronal geometries have been investigated. The simplest and best-studied geometry is the ``lamppost'' model, where the corona is assumed to be  an isotropically-radiating point source located at a few $R_g$ above the BH \citep{Matt91,Reynolds97,Reis13}.
Other coronal geometries that have been explored include an off-axis compact corona or an extended coronal source \citep[see][]{Wilkins12}. Coronae with geometry changing on short timescales, likely associated with changes in the accretion states, have also been reported recently \citep{Kara19}.

While most of the effort described above assumes a thin Keplerian disk structure, recently, some efforts have also been made to incorporate finite disk thickness from analytical disk models \citep{Wu2007, Taylor18} or realistic structures from thin disk simulations \citep[e.g.,][]{Schnittman2013}. However, theoretical investigations on this topic have not been extended to accretion flow with optically-thick winds, which can be formed when a disk undergoes super-Eddington accretion. We expect the X-ray reverberation signatures from super-Eddington disks to be very different to those arising from thin disks, because the coronal emission will no longer be reprocessed by the surface of the (thin) disk but instead by the optically thick winds. Also, we expect the trajectories of the reflected photons and their redshifts to be radically different due to the different geometry and motion of the optically thick winds as compared to the usually assumed equatorial Keplerian flow. One of the main reasons for the lack of literature on this topic is that the detailed structure of super-Eddington accretion flow has not been resolved by simulations until about a decade ago. Another main reason is that observational evidence of X-ray reverberation from a definitively super-Eddington accretion flow has only been reported recently in \citet{Kara16}, where they observed the energy-lag spectrum and a strongly blueshifted Fe line from the jetted TDE, {\it Swift} J1644. In order to successfully explain the salient observational features, the authors also conducted preliminary theoretical modeling by assuming the Fe lines were reflected by a conical funnel moving at 
a speed of few$\times 0.1 \ c$, which corresponds to the speed of winds launched from a super-Eddington accretion disk formed by stellar debris.

As a first step towards better understanding X-ray reverberation from super-Eddington accretion flows and motivated by the framework described in \citet{Kara16}, we study the fluorescent Fe K$\alpha$ emission line profiles derived from state-of-the-art super-Eddington disk simulations. We study in details the profile of a super-Eddington disk, previously simulated in \citet{Dai18}, and calculate the reflection surface of the coronal emission, which is embedded in the optically thick winds, and its properties (Section \ref{sec:Disk_profile}). 
Since the wind region has a much lower density than dense thin disks, the ionization level of the reflection surface is naturally higher, so hotter fluorescent Fe K$\alpha$ lines with higher rest-frame energies are expected to be produced. 
Next, we apply GR ray-tracing techniques to trace the coronal photons from being reflected by this surface to reaching a faraway observer. We calculate the theoretical Fe K$\alpha$ line profile and check its dependence on various parameters such as the observer inclination angle and coronal location (Section \ref{sec:line}). Furthermore, we do a systematic comparison between the Fe K$\alpha$ line profiles produced from the simulated super-Eddington accretion disk and those produced from standard thin disks (Section \ref{sec:compare}). We show that a typical Fe K$\alpha$ line from super-Eddington disks exhibits a large blueshift, while, depending on the observer's inclination angle, the thin disk counterparts are either dominated by redshift or show a much lower blueshift. Also, the Fe line from a super-Eddington disk has a more symmetric shape with respect to the line center.
Therefore, these morphological differences of the Fe K$\alpha$ line can effectively serve as indicators of super-Eddington accretion rates. Lastly, we briefly compare the theoretical predictions to the observation of the Fe K$\alpha$ line from the jetted TDE, \emph{Swift} J1644, and discuss the implications and future directions of this work  (Section \ref{sec:discussion}).

\section{Disk profile, funnel geometry and ionization level} \label{sec:Disk_profile}
\begin{figure}
\centering
\figurenum{1}
\includegraphics[width=\linewidth]{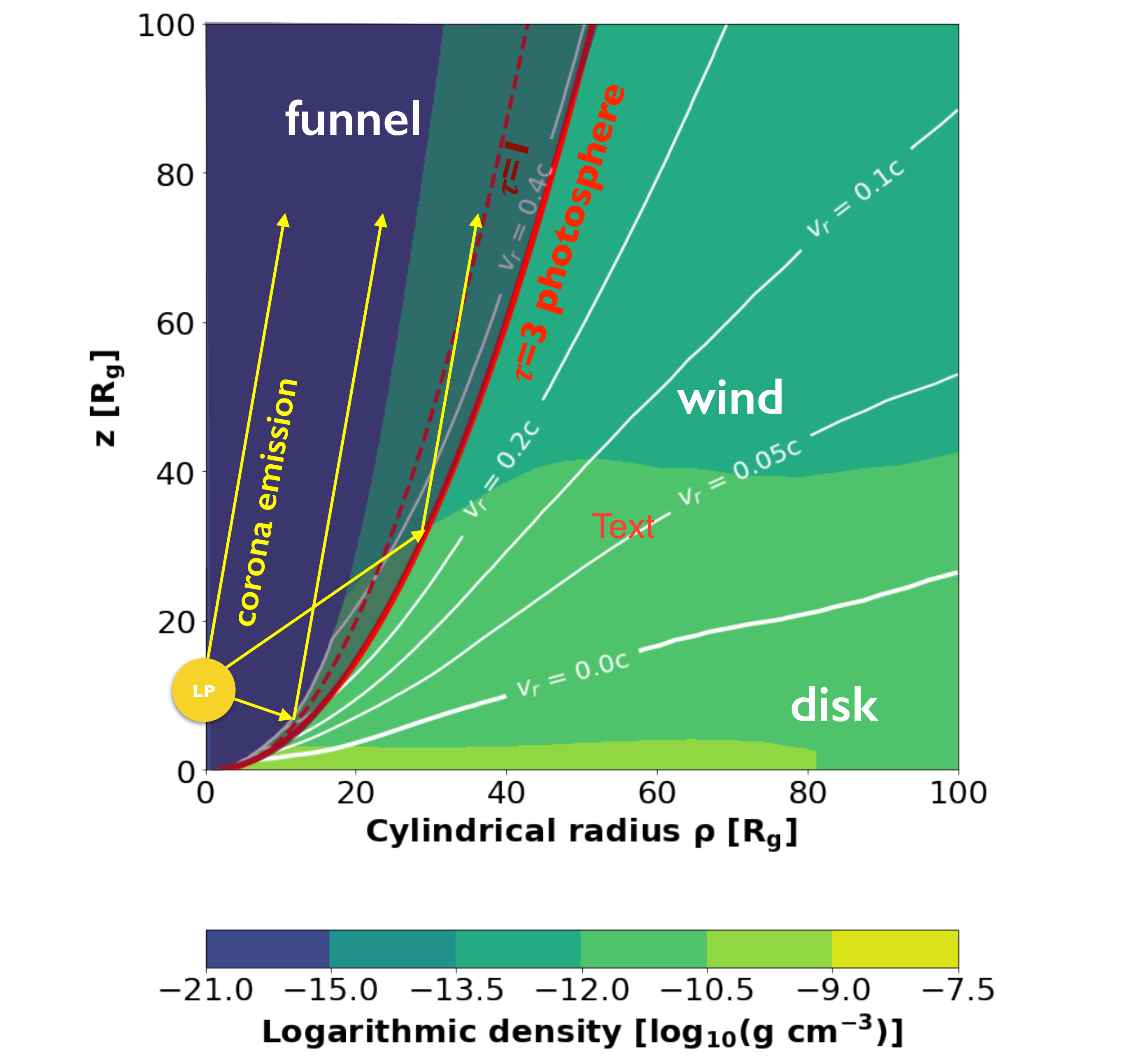} 
\label{fig:result:photosphere}
\caption{
\textbf{The geometry of the disk, winds, corona and its reflection surface for the simulated super-Eddington disk structure.} The density of the disk and winds are shown by the background color in $\log_{10}$ of cgs units [$\rm g \ cm^{-3}$]. Here, the x and y axis are the cylindrical radius $\rho$ and the vertical height $z$, both in units of the gravitational radius $R_g$. We have illustrated the jet (with $\beta<1$) as the dark blue region around the pole. The white lines are contours of constant radial velocity in the wind region, which shows that the winds move faster at small inclination angles. An optically-thin funnel (the shaded region) exists around the pole, which is surrounded by optically-thick winds.
The yellow circle represents the (artificially placed and size-exaggerated) lamppost corona, located at a height of 10 $R_g$ above the BH. The thick red line and thin dashed red line are respectively the electron-scattering photosphere with a Thompson optical depth of $\tau=3$ and $\tau=1$ integrated from the corona, and they represent the reflection surface for coronal emission.
An observer looking down the funnel can see the direct emission from the corona, as well as the coronal emission that is reflected by the funnel (including the Fe K$\alpha$ fluorescent lines).  }
\end{figure}

First, let us briefly summarize the basic properties of super-Eddington accretion disks relevant for X-ray reverberation.
In order to do this, we set the super-Eddington disk, simulated in \citet{Dai18}, in a classical X-ray reflection picture. Fig. \ref{fig:result:photosphere} shows an illustration of the corona, disk, wind and funnel structures. The disk surrounds a supermassive BH with mass of $5\times10^6~M_\odot$ and a fast dimensionless spin parameter of 0.8. We calculate the time and azimuthal-averaged profile of the disk after the inflow equilibrium has been established to $r\approx 100~R_g$ ($R_g=GM_{\rm BH}/c^2$ is the gravitational radius of the BH, with $G$ being the gravitational constant and $M_{\rm BH}$ being the BH mass). The averaged accretion rate onto the BH is about $15~\dot{M}_{\rm Edd}$ during this phase. Consistent with theoretical predictions, the inner disk extends within the ISCO since the thick disk is supported by radiation pressure. A relativistic jet is launched magnetically due to the large-scale magnetic flux threading the inner disk region. Winds, which are denser and slower towards the equator, are also launched from the thick accretion disk, thus creating an optically-thin ``funnel'' near the pole. Here we adopt the standard lamppost model and place a hard X-ray emitting and compact corona at a height of $h_{\rm LP}= {\rm few}\times R_g$ above the BH. (Note that the corona is placed artificially since the GRRMHD disk simulation in \citet{Dai18}, as well as other global accretion disk simulations, cannot self-consistently produce compact coronae.) The emission of the corona is reflected by and irradiate the wall of the funnel.  We have also cut out the jet region (where the gas pressure over the magnetic pressure $\beta\approx 1$) since the gas density in the jet can be numerically boosted due to the large magnetic pressure near the jet base.  

We take a simplified treatment in calculating the coronal reflection surface by assuming that the coronal photons are reflected by a photosphere of a single  electron-scattering optical depth as seen by the corona. The optical depth is calculated by:
\begin{equation} \label{eq:method:tau}
\tau_{\rm es}=\int \kappa_{es}\ \mathrm{d}l=\int 0.2(1+\chi ) \frac{\rm cm^2}{\rm g} dl=\int 0.34 \frac{\rm cm^2}{\rm g} \mathrm{d}l,
\end{equation} 
where $\kappa_{es}$ is the Thomson electron-scattering opacity and $\chi=0.7$ corresponds to the mass fraction of hydrogen with solar abundances. Here, only Newtonian calculations are used for the optical depth without considering the GR light-bending effect. Since the Fe lines are produced within the first few optical depths \citep{Matt93}, we have calculated the $\tau=1$ and $\tau=3$ photospheres and showed them in Fig. \ref{fig:result:photosphere}. The two reflection surfaces lie close to each other since the density of the wind increases steeply away from the pole. Either surface resides mostly in the optically-thick winds with large outflow velocities ($v_r>0.1c$). Also, the reflection surface has a narrow half-opening angle between $10-15^\circ$. Therefore, the optically thick disk and wind will obscure the coronal emission when viewed from side, and the corona and its reflection emission can only be seen by an observer looking directly down the funnel. 

\begin{figure}
\centering
\begin{minipage}{.45\textwidth}
\centering
\figurenum{2a}
\includegraphics[width=\linewidth]{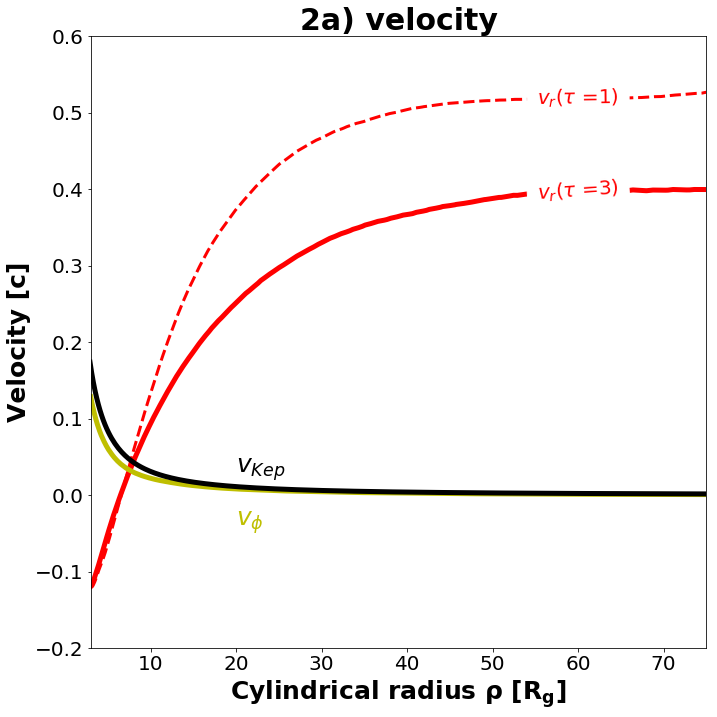}
\label{fig:result:velocity} 
\end{minipage}
\begin{minipage}{0.45\textwidth}
\centering
\figurenum{2b}
\includegraphics[width=\linewidth]{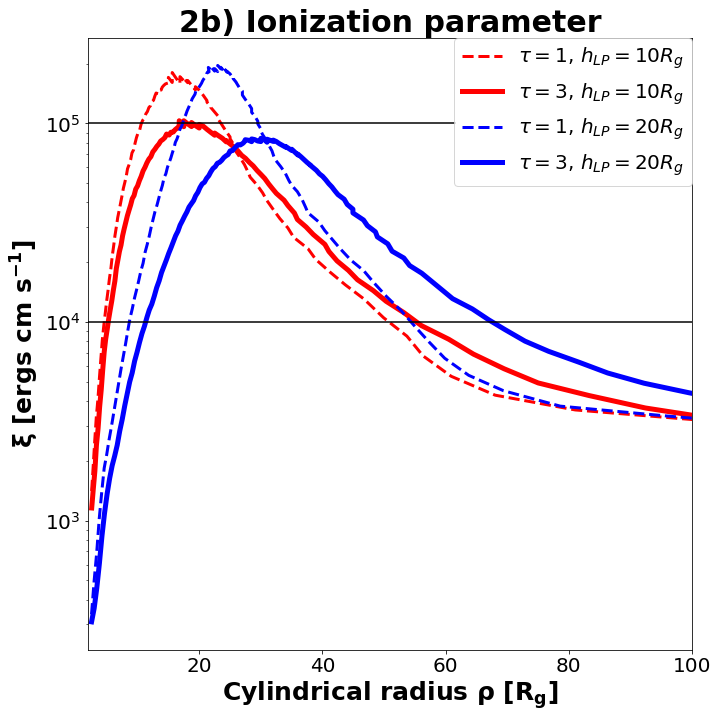} 
\label{fig:result:ionisation}
\end{minipage}
\figurenum{2}
\caption{\textbf{Properties along the reflection surfaces.}
\textbf{2a. The radial and rotational velocity of the reflection surface.} The velocities plotted here are the equivalent Newtonian three-velocity converted from the GR four-velocity of the gas in the simulated accretion flow. The red curve is the radial velocity of the photosphere with $\tau=3$ and the thin dashed curve is that for the $\tau=1$ photosphere. The rotational velocities of the two photospheres lay almost on top of each other (yellow curve). As a comparison, we also plot the Keplerian rotational velocity (black curve), and show that the rotation of the thick disk/wind is sub-Keplerian. $v_\theta$ is relatively unimportant and is therefore not illustrated. Along the reflection surface, it can be seen that the wind starts to accelerate and eventually it saturates at a terminal velocity of $v_{r}\approx$ 0.3-0.5 c. The outward radial motion of the wind dominates over the rotation beyond $\rho \approx 10 \ R_{ g}$. Therefore, the Fe reflection spectrum is primarily determined by the radial motion of the winds instead of the Keplerian motion of the (thin) disk. 
\textbf{2b. The ionization level of the reflection surface.} We show the ionization parameter, $\xi$, in cgs units [$\rm ergs \ cm \ s^{-1}$] along the photosphere as a function of the cylindrical distance $\rho$. The ionization level is not very sensitive to the choice of coronal height or optical depth of the photosphere. With the assumption that the hard X-ray coronal luminosity scales similarly with the accretion rate as in standard thin disks, then for our system we have $L_x = 1\%\ \dot{M}c^2 \approx 0.1~L_{\rm Edd}$, and thus the ionization level is much higher compared to the thin disk case ($\xi \lesssim 1000~\rm{erg~cm~s^{-1}}$). }Therefore, the production of hot Fe lines is favored. The two horizontal black lines mark the range of the ionization parameter where the hot Fe K$\alpha$ fluorescent line with rest-frame energy of 6.97 keV  is likely produced. \label{Fig:photosphere}
\end{figure}

Next, we investigate the motion of the reflection surface.
As shown in Fig. \ref{fig:result:velocity}, close to the BH the photosphere is embedded within a disk inflow region, but beyond $ \rho\approx 6~R_g$ ($\rho$ is the cylindrical distance) it is embedded within the outflows. Along the photosphere, the radial velocity, $v_r$, increases and then saturates towards a terminal speed of $v_r \approx 0.4-0.5 \rm c$. The photosphere is rotating at sub-Keplerian velocity, and the radial velocity dominates over the rotational velocity after $\rho\approx 10\ R_g$. Therefore, one will expect that the X-ray reflection spectrum is dominated by the radial motion of the winds instead of the Keplerian motion of the disk as in the case of a standard thin disk. 
The radial velocity of the photosphere of $\tau=3$ has been fitted (within 5\% deviation in the wind region) with the following equation:
\begin{equation}
  \frac{ v_r}{c}=0.5-\frac{6.2 R_g}{6 R_g+\rho}.
\end{equation}

Last, we investigate a crucial parameter -- the ionization level. The ionization parameter, $\xi$, depends on the density of the reflection surface and the hard X-ray irradiation flux of the corona, and it is calculated as \citep{Reynolds97}:
\begin{equation}
    \xi(r)=\frac{4\pi F_x}{n(r)},
\end{equation}
where $n(r)$ is the gas number density of the reflection surface and $F_x$ is the coronal flux crossing the reflection surface. Using the Newtonian irradiation profile, the coronal irradiation flux can be expressed analytically as:
\begin{equation} \label{eq:Newtonian:irradiation}
  F_x=\frac{L_{x}  \ \mathrm{cos}(\psi)}{4\pi R_{\rm LP}^2},
\end{equation} 
where $L_x$ is the luminosity of the hard X-ray coronal photons, $R_{\rm LP}$ is the distance from the lamppost corona to the the gas element on the reflection surface and $\psi$ is the angle the vector from the corona to the reflection surface produces with the normal vector of the reflection surface.

The ionization level will determine whether Fe K$\alpha$ fluorescent lines can be produced, and if so, whether a higher or lower ionization line (e.g., 6.4, 6.7 or 6.97 keV) is emitted \citep{Ross1999,Ballantyne2001}. Since there is still not a good understanding on how the corona power should scale with the accretion rate in super-Eddington accretion, we assume that the coronal luminosity scales with the accretion rate as $L_x=\eta_x \times \dot{M}c^2$, and we adopt $\eta_x=0.01$ as conventionally used for thin disk coronal models \citep{Reynold99}. Under this assumption, the coronal luminosity is $L_x \approx 0.1 \ L_{\rm Edd}$ for our simulated disk with an accretion rate of about $15\times \dot{M}_{\rm Edd}$.
As illustrated in Fig. \ref{fig:result:ionisation}, the ionization level is several orders of magnitudes higher than that of the theoretical thin disk \citep[usually with $\xi\lesssim 100~\rm{erg~cm~s^{-1}}$ outside the ISCO, see e.g.,][]{Reynolds97}. 
This is mainly because the density of the gas in the wind is much more dilute in comparison to a dense thin accretion disk. 
Fig. \ref{fig:result:ionisation} also shows that the ionization level mostly lies between $10^4$ and $10^5 \ \rm ergs\ cm\ s^{-1}$, independent of the assumptions for the $\tau$-value of the reflection surface or the lamppost height. This high ionization interval strongly favors the production of the hot and warm Fe K$\alpha$ lines with rest-frame energies of $6.97$ and $6.7$ keV instead of the cold $6.4$ keV Fe line, usually assumed to be produced from thin disks \citep{Ballantyne2001}.

\section{Line profile} \label{sec:line}
In this section, we study how an initial Fe K$\alpha$ line emitted at a single energy in the co-moving frame of the emitting gas appears skewed and broadened due to the relative motion and different gravitational potentials between the emitting gas and the observer. This is calculated using general-relativistic ray-tracing. 
In Section \ref{sec:line:methods}, we give a brief summary of the GR ray-tracing theory and introduce our methodology. In Section \ref{sec:line:thick_disk}, we present the calculated Fe K$\alpha$ emission line profiles and show how it depends on physical parameters such as the viewing angle, the choice of the optical depth for the reflection surface and the height of the corona.

\subsection{Methodology for GR Ray-Tracing}\label{sec:line:methods}
When photons are emitted by gas in one frame and observed in another frame, the intensity shift of the photons between the two frames can be calculated using the Louivilles Theorem, which states that the photon phase-space density, $I_{\nu}/{\nu^3}$, is conserved along the ray \citep{Cunnigham75}. Here, $I_{\nu}$ is the specific intensity at frequency $\nu$, and we denote $I_{\nu_o}^{\rm obs}$ to be the observed specific intensity at the observed frequency $\nu_o$ and $I_{\nu_e}^{\rm emit}$ to be the emitted specific intensity at the emitted frequency $\nu_e$. Thus we have: 
\begin{equation} 
I_{\nu_o}^{\rm obs} = I_{\nu_e}^{\rm emit} (\nu_o/{\nu_{\rm e} })^3 = g^{3} I_{\nu_e}^{\rm emit}.
\end{equation}
Here, $g$ is the energy shift (or the redshift) from the emitted photon to the observed photon, and it is defined by:
\begin{equation} \label{eq:method:g-factor}
    g\equiv\frac{\nu_o}{\nu_e}=\frac{E_{\rm obs}}{E_{\rm emit}},
\end{equation}
where $E$ is the energy of the photon. Therefore, we can calculate the total intensity, integrated over all frequencies, by:
\begin{equation}
I^{\rm obs}= \int I_{\nu_o}^{\rm obs} \ d\nu_o 
\propto \int I^{\rm emit}_{\nu_e} g^3 \ d\nu_o 
= \int I^{emit}_{\nu_e} g^3 \ d(g~\nu_{e}) 
\propto g^4~I^{\rm emit}.
\end{equation}

For faraway sources, we point out that the observed flux and intensity are proportional to each other. The reason for this is that the specific flux at the observed frequency $\nu_o$, $F^{\rm obs}_{\nu_0}$, is the observed specific intensity $I_{\nu_0}^{\rm obs}$, integrated across the solid angle of the emitting surface as seen by the observer. Therefore, we have:
\begin{equation}
     F^{\rm obs}_{\nu_o} = \int_{\rm source} I_{\nu_o} ^{\rm obs} \mathrm{cos}(\theta_i) \ d\Omega \approx I_{\nu_o}^{\rm obs} \Omega \propto I_{\nu_o}^{\rm obs}.
\end{equation}
Here $\Omega$ is the solid angle of the reflection surface as seen by the the faraway observer, and $ \theta_i$ is the angle between the direction of the BH and incoming rays from the reflection surface. Since the observer is assumed to be far away, then $\mathrm{cos}(\theta_i)\approx 1$. Therefore, the total flux goes like $F\propto g^4$ as well.

We employ a GR ray-tracing code \citep[][based on equations from \citealt{Fuerst04}]{Dai2012} to calculate the photon trajectory from the emitting gas to a faraway stationary observer. The code uses the Boyer-Lindquist spherical coordinate system to describe the space-time around a Kerr black hole with the line element (in geometric units $G=M=c=1$):
\begin{equation}
ds^2 =-\Bigg(1-\frac{2r}{\Sigma}\Bigg)~dt^2 - \frac{4ar  \mathrm{sin^2}\theta}{\Sigma}~dt~d\phi + \frac{\Sigma}{\Delta}~dr^2 + \Sigma~d\theta^2 + \Bigg(r^2 + a^2 + \frac{2a^2r \mathrm{sin^2}\theta}{\Sigma}\Bigg) \mathrm{sin^2}\theta ~d\phi^2,
\end{equation}
where ($t, r,\theta, \phi$) is the Boyer-Lindquist spherical coordinates and $\Delta=r^2-2\ r+a^2$ and $\Sigma=r^2+a^2\ \mathrm{cos^2}\theta$. 
In this code, we calculate the geodesic of particles by evolving the six variables $t, r, \theta, \phi, p_r, p_\theta$ using the following equations:
\begin{align}
    & p_t = -E        &&  \dot{t} = E + \frac{2 r (r^2+a^2)E-2 a L}{\Sigma \Delta} \nonumber \\
    & p_r = \frac{\Sigma}{\Delta} \dot{r}   &  &\dot{p}_r = \frac{(r-1)\big((r^2+a^2)H- \kappa \big) + r H  \Delta + 2 r(r^2 + a^2)E^2-2 a E L}{\Sigma  \Delta} - \frac{2 p_r ^ 2(r-1)}{\Sigma} \nonumber\\
    & p_\theta = \Sigma \dot{\theta}  &    &\dot{p}_\theta = \frac{\rm sin\theta  \rm cos\theta}{\Sigma } \Bigg(\frac{L^2}{\rm sin^4\theta}- a^2(E^2+H) \Bigg) \nonumber\\
    & p_\phi = L && \dot{\phi} = \frac{2 a r E + (\Sigma-2 r) L/ \mathrm{sin^2}\theta}{\Sigma  \Delta} \nonumber\\
\end{align}
Here, $p_t$ and $p_\phi$ are constants of motion representing the conservation of energy, $E$, and angular momentum around the spin axis, $L$. Furthermore, we have two additional constants: $H$ is the Hamiltonian (which is 0 for photons and -1 for massive test particles) and $\kappa=Q+L^2+ a^2(E^2+H)$, where $Q$ is the Carter's constant given by $ Q=p_\theta^2 -a^2E\mathrm{cos^2}\theta + L^2\mathrm{cot^2}\theta$.

Evolving these six variables along a ray enables one to calculate the photon energy shift, $g$, from the frame of the emitting gas to the faraway observer as:
\begin{equation}
    g=\frac{E_{\rm obs}}{E_{\rm emit}}=\frac{(p_{\mu} u^{\mu})_{\rm obs}}{(p_{\mu}  u^\mu)_{\rm emit}}=\frac{-E}{(p_{\mu}  u^{\mu})_{\rm emit}}.
\end{equation}
The four-velocity for a faraway stationary observer only has a time component of $u^t=\big( 1-2r/\Sigma \big)^{-1} \approx 1$, and $u^{\mu}_{\rm emit}$ is the four-velocity of the emitting gas along the reflection surface. For a (prograde) thin disk, the reflection surface lies in the equatorial plane and orbits with  the GR Keplerian four-velocity \citep{Bardeen72}:
\begin{equation}
     \tilde{u}_{\rm disk}=u^t(1,0,0,\Omega), \ \ \ {\rm where} \ \ \Omega=(a + \rho^{3/2})^{-1}, \ \ \ {\rm and} \ \ u^t=\frac{a+\rho^{3/2}}{\sqrt{\rho^3-3\rho^2+2 a \rho^{3/2}}}.
\end{equation}
When calculating the Fe line profile from the super-Eddington disk, the four-velocity of the emitting gas element along the chosen photopshere is directly obtained from the GRRMHD simulation. Since the irradiation parameter drops very fast with distance from the BH, we only include emission from the disk region extending from the ISCO to $100 \ R_g$ when calculating the thin disk profile. For super-Eddington disks we include out to cylindrical radius of $\rho = 100 \ R_g$. 

We also assume that the emitted fluorescent line intensity scales with the coronal irradiation flux crossing the reflection surface (see Eq. \ref{eq:Newtonian:irradiation}). Therefore, the observed Fe K$\alpha$ line intensity can be calculated using the following equation:
\begin{equation} \label{eq:method:observed}
 I^{\rm obs} = I^{\rm emit} \ g^4\propto g^4 \  \frac{\mathrm{cos}(\psi)}{R_{\rm LP}^2}.
\end{equation}
In the ray-tracing code, we use the standard fifth-order Runge-Kutta integrator to evolve these equations together with adaptive step sizes, so the curved photon trajectories close to the BH can be accurately traced.
We also adopt the standard procedure to start from an image plane, placed at $r=1000 \ R_g$, and trace the photons backwards in time to the accretion disk to save computing power.

\begin{figure}
\centering
\figurenum{3}
\begin{minipage}{0.45\textwidth}
\centering
\figurenum{3a}
\includegraphics[width=\linewidth]{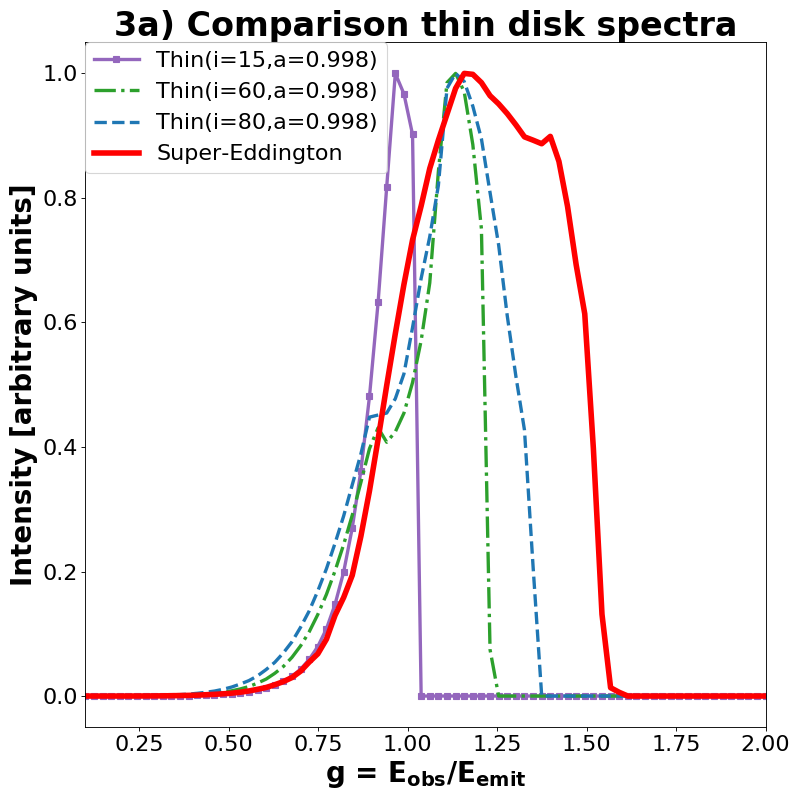}
\label{fig:result:comparison_thick_thin}
\figurenum{3c}
\includegraphics[width=\linewidth]{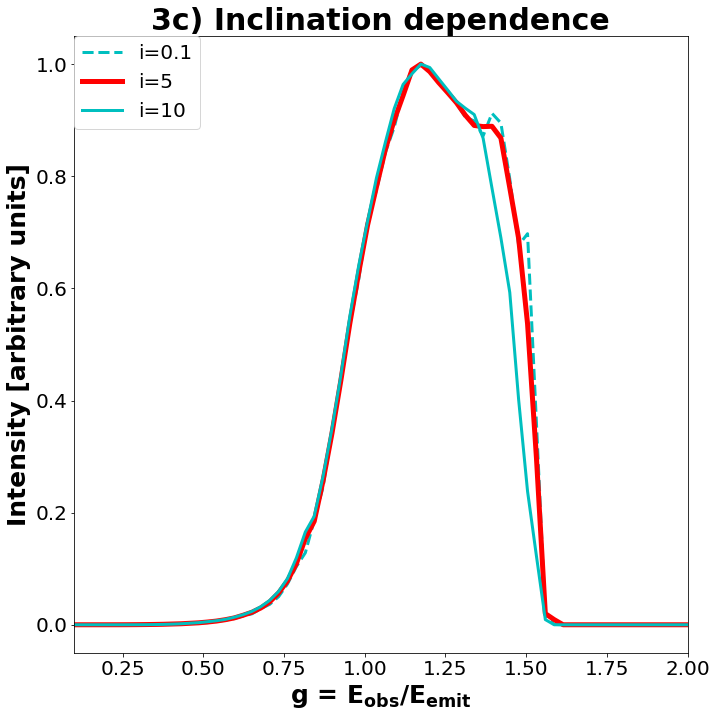}
\label{fig:result:comparison_inclination}
\end{minipage}
\begin{minipage}{0.45\textwidth}
\centering
\figurenum{3b}
\includegraphics[width=\linewidth]{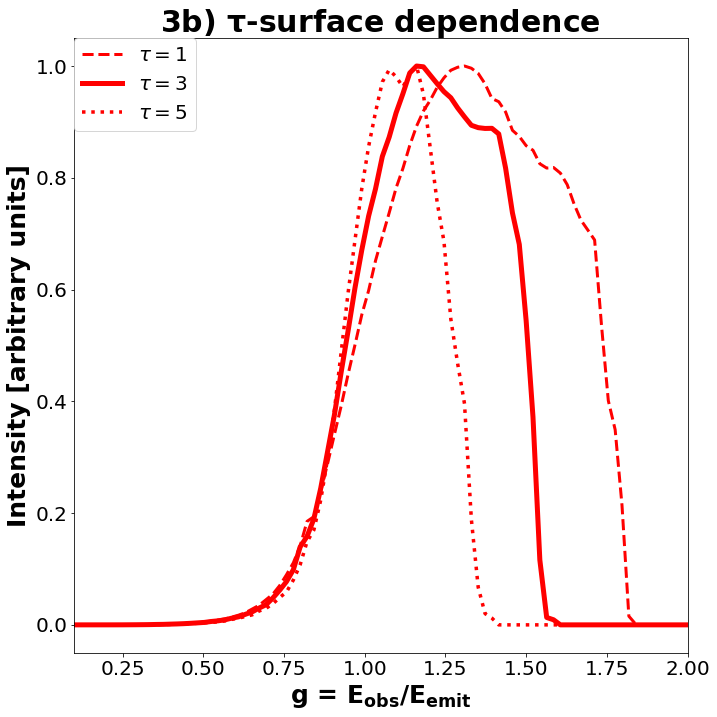}
\label{fig:result:comparison_tau}
\figurenum{3d}
\includegraphics[width=\linewidth]{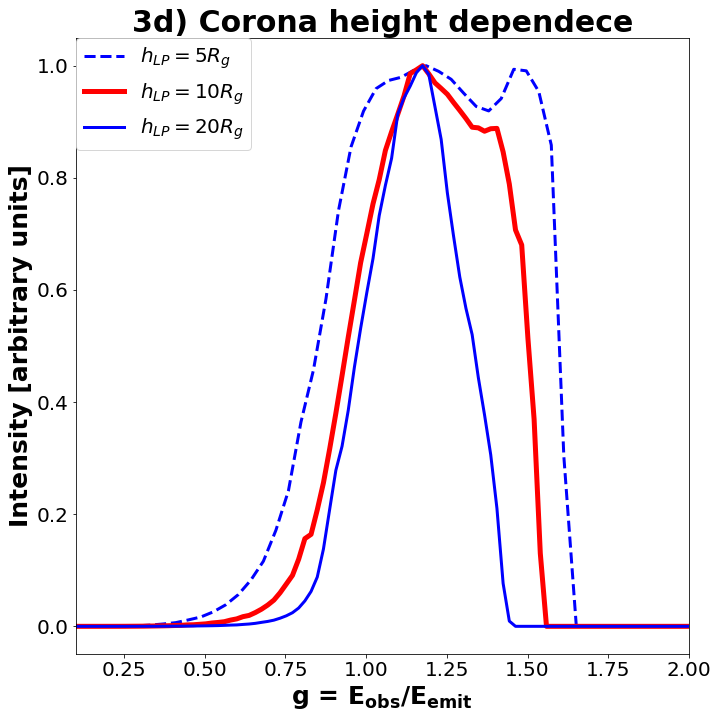} 
\label{fig:result:comparison_LP}
\end{minipage}
\caption{\textbf{The intensity of fluorescent Fe K$\boldsymbol{\alpha}$ line profiles from the simulated super-Eddington disk with different $\boldsymbol{\tau}$-surfaces, inclination angles and lamppost heights, and also in comparison with thin disk line profiles.} The thick red line in all panels represent the fiducial super-Eddington disk case using $\tau=3$, $h_{\rm LP} = 10 \ R_g$ and $i=5^\circ$. All the line profiles are scaled to have the maximum at 1.
\textbf{3a. Comparison between the Fe line from the super-Eddington disk and representative thin disks.} The thin disk Fe lines are plotted using thin lines. All the thin disks have the same spin parameter of $a=0.998$. It can easily be seen that the Fe line profile of the face-on super-Eddington disk is even more blueshifted than line profiles of thin disks viewed from along the disk direction.
\textbf{3b. Dependence on the choice of the $\boldsymbol{\tau}$-surface.} We fix the height of the corona at $10 \ R_g$ and the inclination at $5^\circ$, but vary the optical-depth of the reflection surface, $\tau$. With increasing $\tau$ values, the funnel becomes wider with a lower wind terminal velocity along the funnel wall, which reduces the overall blueshift and the width of the Fe line produced.
\textbf{3c. Dependence on the viewing angle from the pole.} Here, we fix $h_{\rm LP} = 10 \ R_g$ and $\tau=3$ and vary the viewing angle to the observer. Since the half-opening angle of the funnel is very narrow ($i \approx 10-15 ^\circ$), there is not much freedom in $i$ for the observer to see the coronal emission and reflection. Therefore, the Fe line profiles are almost independent of $i$ as long the observer is looking down into the funnel.
\textbf{3d. Dependence on the height of the compact corona.} We only vary $h_{\rm LP}$ and keep everything else the same as in the fiducial case. A lower corona gives more weight to the irradiation of the inner disk and therefore induces more gravitational redshift. It also induces more blueshift to the line profile since the funnel, as seen by the corona, becomes narrower and lies in a faster wind region. Therefore, the width of the Fe line is correlated with the height of the corona.}
\label{fig:result:compare}
\end{figure}

\subsection{Fe K$\alpha$ Reflection Spectrum from Super-Eddington Accretion Flow} \label{sec:line:thick_disk}
In this section, we will present the calculated X-ray fluorescent Fe K$\alpha$ line profiles from the simulated super-Eddington accretion disk. These line profiles look very different from the thin disk Fe lines primarily because the coronal reflection surface has fundamentally different geometry and motion. While various simulations of super-Eddington accretion disks have shown that the details of the wind profile can depend on the BH mass, spin or the accretion rates \citep{Mckinney15, Sadowski16,Jiang17}, these simulations have also shown that the basic structure of the accretion inflow and outflow stays the same and the wind speeds in the funnel region are consistent to the first order. Therefore, the basic coronal irradiation picture, as illustrated in Fig.\ref{fig:result:photosphere}, should generically apply for super-Eddington systems. In this work, we only calculate the Fe reflection spectrum from one simulated super-Eddington disk structure \citep{Dai18} in order to highlight its morphological differences to the standard Fe line spectra produced by thin disks, since the exploration on how the exact funnel geometry or wind speed depends on various physical parameters goes beyond the scope of this short letter.

Our fiducial configuration for the Fe line profile of the super-Eddington accretor is obtained using the following parameters: $\tau=3$ \cite[following][]{Matt93}, $h_{\rm LP} = 10 \ R_g$ and $i=5^\circ$ (from the pole), and it is plotted in all panels of Fig. \ref{fig:result:compare} using thick red lines.
We plot this line profile together with a few representative Fe line profiles of thin disks in Fig. \ref{fig:result:comparison_thick_thin}. Consistent with previous studies \citep{Reynolds97, Dabrowski97, Reynold99}, the thin disk Fe line profile has a strong viewing-angle dependence, and it is always skewed to the right due to an extended red wing caused by gravitational redshift. Only when a thin disk is viewed almost edge-on, one can observe a large blueshifted component (with $ g=E_{\rm obs}/E_{\rm emit}>1$) due to the Doppler motion of the disk. However, for the face-on super-Eddington disk case, one can see that the Fe line is very strongly blueshifted because the reflecting gas elements are in the fast winds, which are moving towards the observer looking down the funnel. We will discuss more details of these differences in Section \ref{sec:compare}. 

In the following part of this section, we will check whether the Fe line profiles, produced by the super-Eddington disk,  stay robust against changes in the choice of the optical depth of the reflection surface, $\tau$, and the height of the lamppost corona, $h_{\rm LP}$, and whether they sensitively depend on the inclination angle of the observer, $i$.

First, we investigate the dependence on the choice of photosphere. As shown in Fig. \ref{fig:result:comparison_tau}, the Fe line spectra produced from different $\tau$-surfaces have similar line shapes, while the peak of the Fe line profile is more blueshifted in the case of smaller $\tau$ values.
The reason for this is that the photosphere with a smaller $\tau$ value lies closer to the pole (see Fig. \ref{fig:result:photosphere}) where the wind moves faster. Also, the Fe line profile from the photosphere with a smaller $\tau$ value has a broader width due to the larger velocity variance of the wind speed along this photosphere. Note, we have simplified the reflection geometry to be a reflection surface of a single $\tau$ value, although photons should have gone through multiple scatterings in the optically-thick wind region at different optical depths before being reflected to the observer. A detailed radiative transfer calculation is needed to obtain an accurate Fe line profile. However, the similarities between the line shapes across different $\tau$-surfaces indicate that the Fe line profile obtained this way is a good approximation.

Next, we show in Fig. \ref{fig:result:comparison_inclination} that the observed Fe line profile has a very weak dependence on the inclination angle of the observer -- as long the observer is looking directly into the funnel. Since the funnel is narrow (with a half-opening angle of $10-15^\circ$), so only small changes in the viewing angles are possible, then one would not expect to see any large differences in the Fe line profiles. However, observing the corona and its reflected emission does depend heavily on the viewing angle, in the sense that the optically thick disk and wind will obscure the emission for observers looking at the system from other directions.
The viewing-angle dependence is further weakened since the radial motion of the wind dominates over its rotation.

Last, we investigate the dependence that the height of the corona has on the line profile. Since the photosphere is calculated from the corona, its geometry, velocity and irradiation profile all depend on the location of the corona. In the thin disk case, a compact corona at $h_{ \rm LP}=$ few $R_g$ usually fits the observations \citep[see][]{Reis13}. For a super-Eddington disk, we assume that the corona is located at a similar height, and here we show the Fe line profiles using three different coronal heights of 5, 10, and $20~R_g$ in Fig. \ref{fig:result:comparison_LP}. The photosphere calculated from a higher coronal height has a larger half-opening angle at large radii. Therefore, the corresponding Fe line profile is less blueshifted and narrower. It can further be seen that lowering the height of the corona to $5\  R_g$ will greatly broaden the red wing of the line. This is because the photosphere close to the BH, where the gravitational redshift is the strongest, will receive more irradiating flux and contribute more to the line profile. This effect will become even more dramatic if the irradiation profile is calculated using general relativity with light bending effects.

\begin{figure}
\centering
\figurenum{4}
\begin{minipage}{.45\textwidth}
\figurenum{4a}
\includegraphics[width=\linewidth]{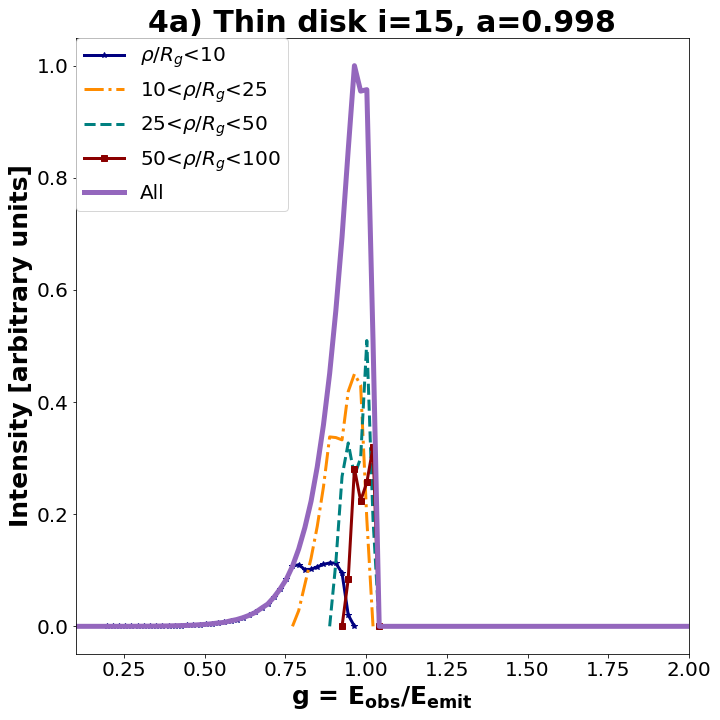} 
\label{fig:result:radius_thin15}
\figurenum{4b}
\includegraphics[width=\linewidth]{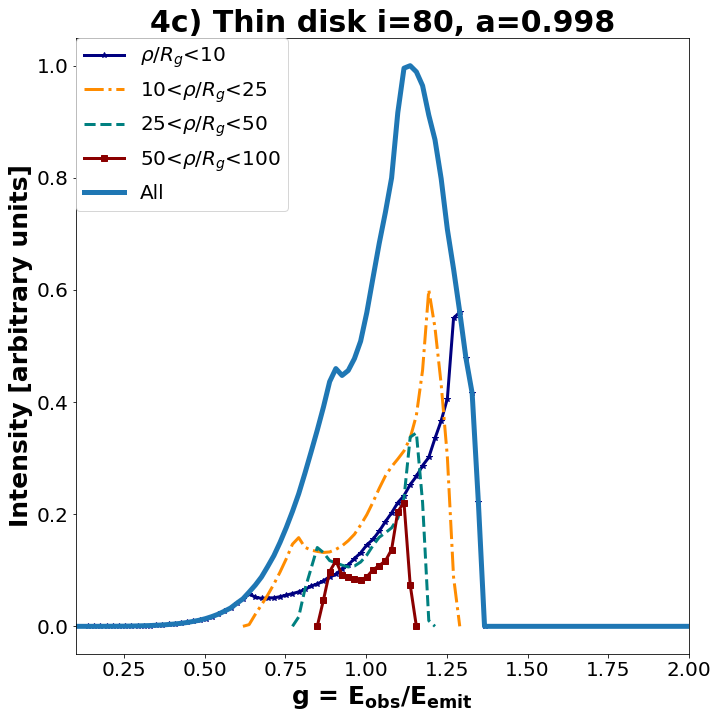} 
\label{fig:result:radius_thin60}
\end{minipage}
\begin{minipage}{0.45\textwidth}
\figurenum{4c}
\includegraphics[width=\linewidth]{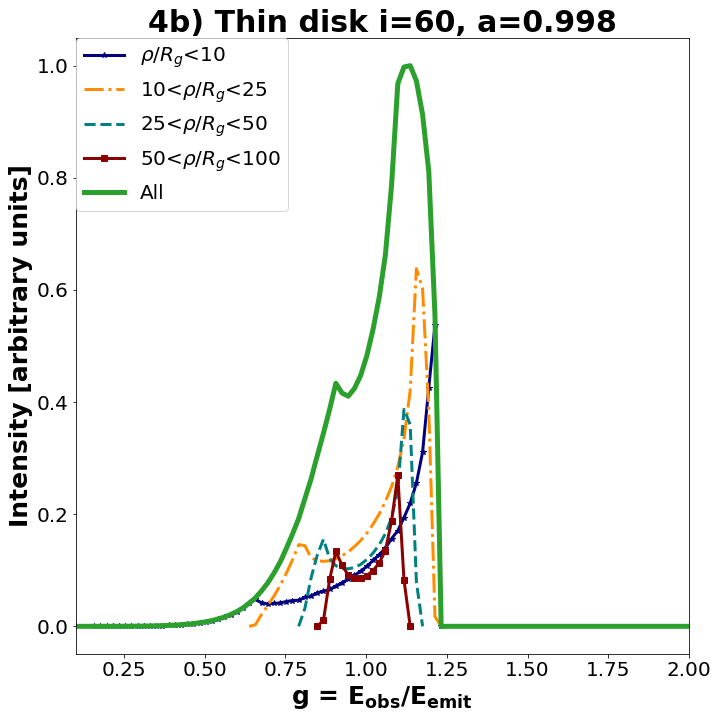} 
\label{fig:result:radius_thin80}
\figurenum{4d}
\includegraphics[width=\linewidth]{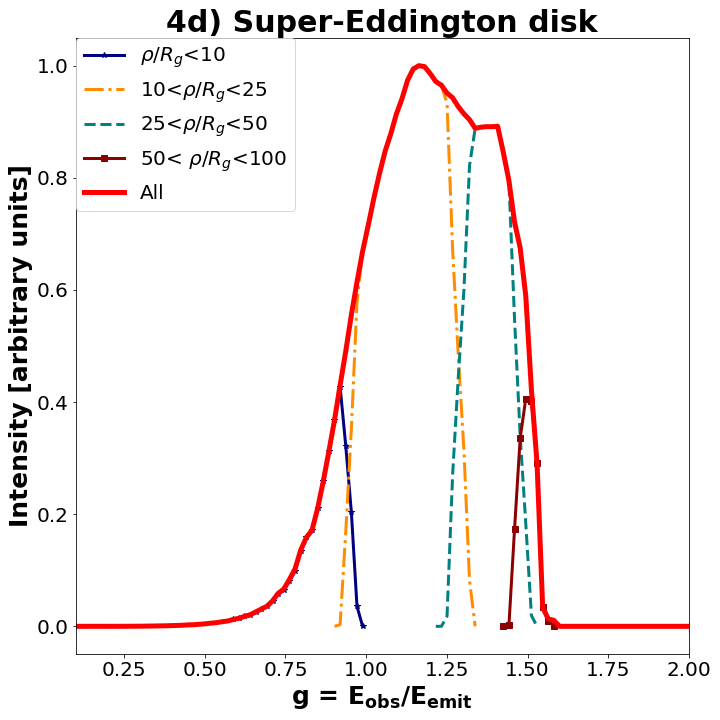} 
\label{fig:result:radius_thick}
\end{minipage}
\caption{\textbf{Decomposition of the Fe line energy profile based on the radii of emission.} All panels show the contribution from each cylindrical radial bin (see their respective labels for interval) towards the total line profile, which is illustrated as the thick solid lines. All line profiles are scaled to have the maximum at 1.
\textbf{4a, 4b, 4c) Radial contribution towards the line profile for thin accretion disks with spin $\mathbf{a=0.998}$ and respective inclination angles of $\mathbf{i= 15^\circ}$, $\mathbf{60^\circ}$ or $\mathbf{80^\circ}$.}
For all three thin disk spectra, the contribution from within $\rho<25 \ R_g$ dominates the total line flux since the irradiation flux drops very quickly with radii. Also, when compared to the line spectra from the outer disk regions, the line spectrum produced from the innermost disk region is the broadest with both the strongest gravitational redshift and Doppler shift (when viewed from side).
\textbf{4d) Radial contribution towards the line profile for the super-Eddington accretion disk.}
It can be seen that the innermost region close to the BH horizon contributes solely to the red wing of the line profile, since the innermost region is dominated by the strong gravitational redshift as the photosphere lies in either inflow or low-velocity outflow regions. It can further be seen that the bluest part of the line profile is produced at the largest distances due to wind acceleration and large terminal velocity. Also, the contribution from $\rho>25\ R_g$ is more significant compared to the thin disk line profiles due to the curvature of the photosphere and relativistic Doppler boosting.} 
\label{fig:result:radius}
\end{figure}

\section{Identifying super-Eddington systems from their Iron K$\alpha$ line profiles} \label{sec:compare}
In the previous section, we have compared Fe line profiles produced from the super-Eddington disk with a few representative produced from (razor) thin disks.
In this section, we will do a further systematic comparison, and show that the Fe line signatures between the two types of disks are morphologically different. 

\textbf{1. Rest-frame energy of the line:} As we have shown in Section \ref{sec:Disk_profile}, the reflection surface in the case of a super-Eddington disk has a much higher ionization level, which will more likely produce hot Fe K$\alpha$ lines, such as the 6.97 keV rest-frame energy line, instead of the cold 6.4 keV line. Apart from this, the shape and the temporal signatures of the line profiles from the two types of disks are also distinguishable, and it is caused by the different geometry in the reflection surfaces together with relativistic effects. In order to demonstrate these difference, we show all the Fe line spectra as a function of the redshift, $g$, instead of the observed energy.

\textbf{2. Photon energy and emission radius correlation:} Photons emitted from various radii on the reflection surface from the two types of disks contribute very differently to the line spectra. For thin disks in Fig. \ref{fig:result:radius_thin15}, \ref{fig:result:radius_thin60} and \ref{fig:result:radius_thin80}, one can see that more than half of the line flux is contributed by photons emitted from within $\rho = 25 \ R_g$ of the disk, since the irradiation profile drops quickly faraway from the BH. 
As a comparison, for the super-Eddington disk case (Fig. \ref{fig:result:radius_thick}), there is a considerable contribution to the line flux by photons emitted from $\rho\approx 25-50 \ R_g$ (which corresponds to a larger radial distance from the BH compared to the $\rho\approx 25-50 \ R_g$ region on a thin equatorial disk). This result can be attributed two factors. First, in the super-Eddington disk case, the irradiation flux drops more slowly at intermediate to large distances from the BH because the reflection surface curves towards the pole. Second, the wind accelerates along the reflection surface, and the photon intensity scales with the blueshift as $I\propto g^4$. Note that for a face-on thin disk, all parts of the disk only produce redshifted photons. Due to the Doppler effects, thin disks with moderate-to-large viewing angles (Fig. \ref{fig:result:radius_thin60} and \ref{fig:result:radius_thin80}) produce a double-peak line feature for each annulus of the disk. Also, the most dramatic redshifted and blueshifted part of the line profile are both produced from the innermost part of the disk ($\rho<10 \ R_g$), where the rotation speed is the fastest and the gravitational redshift is the strongest. However, for a face-on super-Eddington disk, the Fe line photons emitted from the inner part of the reflection surface only contributes to the red wing, and then due to wind acceleration the photon blueshift gradually increases as it is emitted farther away from the BH. Therefore, the case of thin disks and super-Eddington disks should be distinguishable by conducting a careful temporal study on the response at different energy ranges of the Fe line spectra to changes in the coronal continuum emission.

\begin{figure}
\begin{minipage}[b]{0.28\textwidth}
\includegraphics[width=\linewidth]{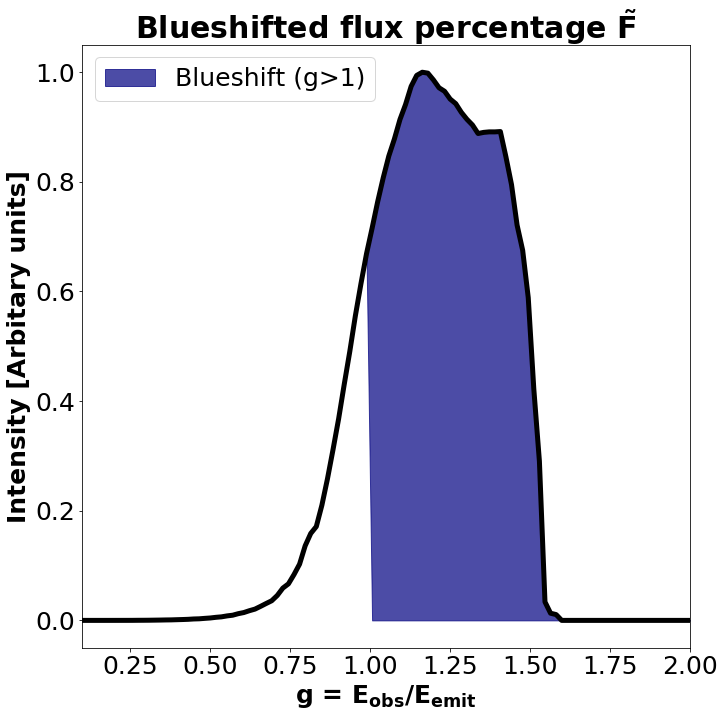}\\
\includegraphics[width=\linewidth]{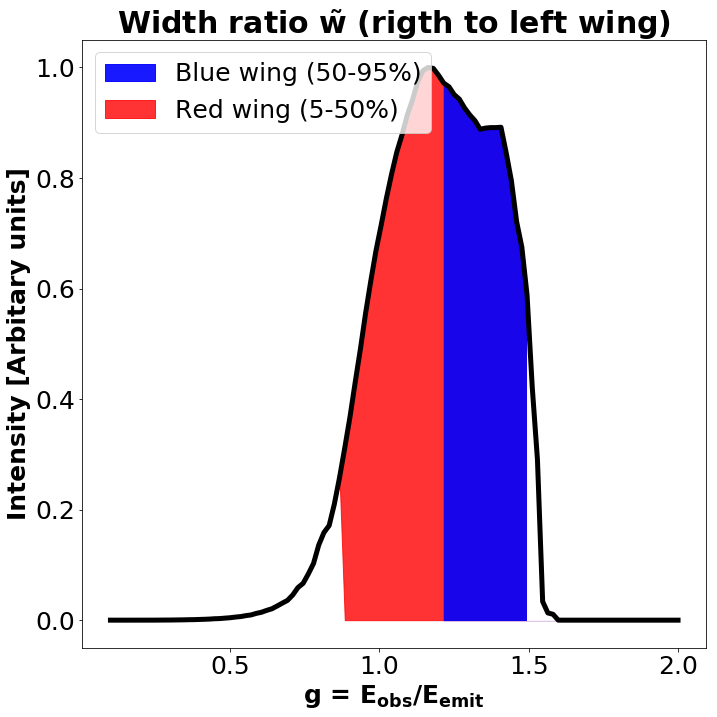}
\end{minipage}
\begin{minipage}[b]{0.6\textwidth}
\includegraphics[width=\linewidth]{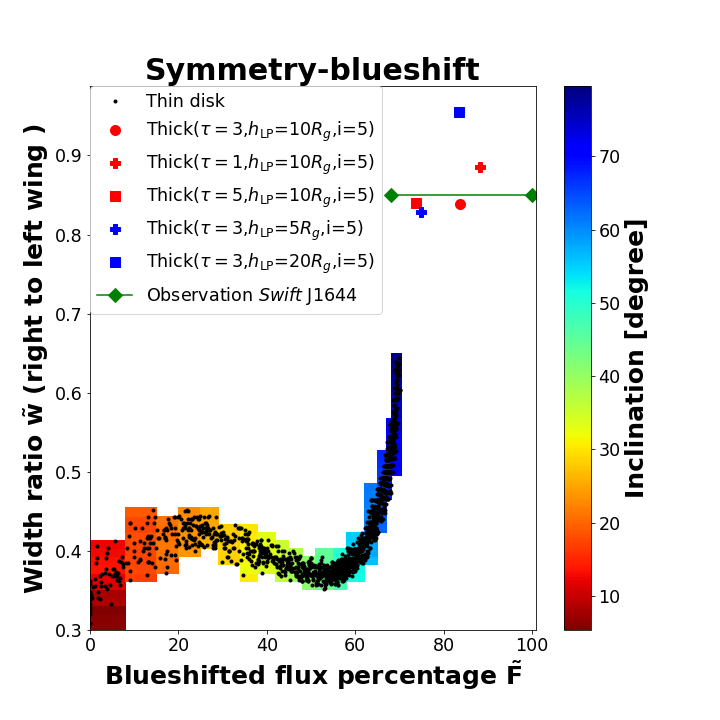}
\end{minipage}
\centering
\figurenum{5}
\label{fig:result:symmetry}
\caption{\textbf{Phase-space diagram illustrating the morphological differences (in terms of the blueshift of the line and the symmetry of the line shape) between Fe K$\alpha$ line profiles from super-Eddington and thin disks.}
The two panels on the left-hand side illustrate how we calculate the two phase-space parameters: \textbf{the blueshifted flux percentage ($\tilde{F}$)} and \textbf{the width ratio ($\tilde{w}$)}.
Here, $\tilde{F}$ is defined as the percentage of the total integrated Fe line flux with observed energy above the rest-frame energy (or $g>1$), and $\tilde{w}$ is the width ratio between the blue wing (where 50\%-95\% of the total line flux lies) to the left wing (with 5\%-50\% of the total line flux).
In the main panel on the right-hand side, we show the phase-space diagram that illustrates these two line profile parameters calculated from both the simulated super-Eddington disk as well as approximately $1500$ thin disks, which we randomly sampled from a uniform distribution of spin parameters between 0 and 0.998 and a uniform distribution of solid angles up to an inclination angle of $80^\circ$. In the phase space diagram, the Fe lines produced by the super-Eddington disk with different choices of $\tau$ and $h_{\rm LP}$ are marked with the large red/blue symbols on the top right corner.  The thin disk lines are marked with the black points, and the color scheme in their background indicates the inclination angle, $i$, of the observer (red -- face on, blue -- edge on). One can see that $\tilde{F}$ is strongly correlated with $i$ for thin disks. However, even the thin disks observed from the largest inclination angles do not produce lines as blueshifted as those from the super-Eddington disk. 
Furthermore, for thin disks, $\tilde{w}$ is always smaller than 1 since their line profiles are always skewed towards the blue side due to the gravitational redshift (which produces an extended red wing) and the relativistic beaming (which breaks the symmetry of the double-peak features). The Fe lines from super-Eddington disks are much more symmetric in shape since emission from the inner disk carry less weight due to the funnel geometry and the large terminal wind speed, which greatly broadens the blue wing of the Fe line spectrum.
The Fe lines observed from the jetted TDE, \emph{Swift} J1644, are marked with two green diamonds. (See Section \ref{sec:discussion} on why there are two data points.) Despite the uncertainty of the line profile as indicated by the green line connecting the two points, the Fe K$\alpha$ line observed from \emph{Swift} J1644 clearly resides in the phase-space region predicted for super-Eddington accreting systems. }
\label{fig:result:phasespace_all}
\end{figure}

\textbf{Blueshift and shape of the line:} When studying Fig. \ref{fig:result:comparison_thick_thin}, one can see that the Fe line profile produced from the super-Eddington disk is both more blueshifted and symmetric with respect to the line center than in any of the thin disk cases. In order to test if these two differences are generic, we have generated a phase-space plot in Fig. \ref{fig:result:phasespace_all}, where we have calculated these two line features for both the super-Eddington disk and a sample of 1500 thin disks. The thin disk sample is created by randomly sampling from a uniform distribution of BH spin between a Schwarzchild BH ($a=0$) to a maximally rotating astrophysical BH ($a=0.998$) \citep{Thorne74}, and at the same time random sampling from a uniform distribution of solid angles with inclination angles up to 80$^\circ$ (beyond which the disk is usually obscured by itself or some torus at large radius).
We quantity the blueshift of the Fe line, $\tilde{F}$, as the percentage of the total integrated line intensity above $g=1$. For the symmetry with respect to the line center, $\tilde{w}$, we define it as the ratio between the width of the blue wing to the width of the red wing. Here, the line center is defined as the place from where each side has half of the integrated line flux. The blue wing refers to the side bluer than the line center containing $50-95\%$ of the line flux, and the red wing is the redder side containing $5-50 \%$ of the total line flux. 
Both these parameters can be visualized in the two panels on the left-hand side in  Fig. \ref{fig:result:phasespace_all}. The main panel on the right-hand side shows the two parameters, $\tilde{F}$ and $\tilde{w}$, calculated from the Fe line profiles of 1500 thin disks and the super-Eddington disk. It can be seen that the Fe line profiles from the super-Eddington disk have a much larger blueshift than all of the thin disk cases. Additionally, for thin disks, a strong correlation between the blueshift parameter, $\tilde{F}$, and the inclination angle of the observer can be seen, and depending on the angle the lines can either be redshifted or blueshifted. (Here, we acknowledge that our Newtonian emissivity profile (Eq. \ref{eq:Newtonian:irradiation}) does not include light bending towards the BH, thus giving less weight to the innermost disk region. Therefore, the difference brought by the spin is underestimated and difficult to observe.) 
We can also see that the Fe line from the super-Eddington disk is significantly more symmetric (with width ratio $\tilde{w} \approx 1$) when compared to the thin disk cases. All the Fe lines produced by thin disks are skewed to the blue wing because of relativistic beaming effects and gravitational redshift. Therefore, the shape of Fe line of a thin disk is generally less symmetric than the super-Eddington case. For the super-Eddington case, we have tested the robustness against the choice of lamppost height ($h_{\rm LP}=5-20 \ R_g$) and optical depth of the reflection surface ($\tau=1-5$). It can be seen that the assumptions on these parameters do not affect the qualitative conclusion.

In summary, we have demonstrated that the super-Eddington disk Fe lines should occupy a significantly different parameter space when analyzing their rest-frame energy, blueshift and the symmetry of the lines profile. When combining these factors, and aided with energy-lag analysis, one can distinguish super-Eddington accretion disks from thin disks.

\section{Discussion and future work} \label{sec:discussion}
In this letter, we have presented a first theoretical investigation of the Fe K$\alpha$ fluorescent line profile produced when a realistic super-Eddington accretion disk, resolved in GRRMHD simulations, is irradiated by a lamppost corona. The main results are summarized below.
\begin{itemize}
\item{The reflection surface of the coronal emission, where the Fe K$\alpha$ fluorescent lines are produced, can be represented by the wall of an optically-thin funnel surrounded by optically-thick winds. The geometry and motion of the reflection surface are therefore very different from the standard case where reflection happens on the surface of a (razor) thin Keplerian disk.}

\item{When viewed at large inclination angles away from the pole, the optically-thick disk and wind produced in super-Eddington accretion will obscure the coronal emission and its reflection spectrum. Therefore, the Fe K$\alpha$ lines can only be seen by observers looking at the system along the funnel direction.}

\item{The ionization level of the reflection surface is high, so the production of hotter Fe K$\alpha$ lines with rest-frame energies of 6.97 or 6.7 keV is favored over the cold 6.4 keV Fe line.}

\item{The Fe line profile from a super-Eddington disk is primarily determined by the geometry of the funnel wall and the motion of the winds. The line spectra is largely blueshifted. Also, photons emitted outside $10~R_g$ contribute more to the line flux due to the acceleration of the wind until large radii and the curvature of the reflection surface.}

\item{The Fe line profile is robust against choosing different values of the optical depth to calculate the reflection surface, which indicates that our simple treatment of assuming that all coronal emission is reflected at a single $\tau$-value surface is a good approximation of the realistic multiple-scattering scenario. The first-order features of the line profile are also robust against changes in the height of the lamppost corona.}

\item{By comparing the Fe K$\alpha$ line profiles from the super-Eddington disk to those from thousands of thin disks sampled with random spin parameters and orientation, we have shown that the Fe K$\alpha$ line profile from a super-Eddington disk is distinctively more blueshifted and symmetric in shape with respect to the line center. This result is also robust against the choice of the lamppost height and the reflection surface $\tau$-value.}
\end{itemize}

Actually, the strongly blueshifted hot Fe K$\alpha$ fluorescent lines from super-Eddington disks, as modeled here, have been observed from the jetted TDE, \emph{Swift} J1644 \citep{Kara16}. TDEs are believed to be super-Eddington accretors if the BH mass is smaller than a few$\times10^7~M_\odot$ \citep{Rees1988, Evans1989, Ramirez-Ruiz09, Guillochon2013, Dai15, Wu2018}, and the existence of a relativistic jet in \emph{Swift} J1644 \citep{Burrows2011, Bloom11, Berger12,2012ApJ...760..103D} further supports that a transient magnetized thick disk has been formed by the stellar debris \citep{Tchekhovskoy14}. 
In \citet{Kara16}, the observed Fe line has an energy peaked at $\approx 8$ keV, which for an Fe K$\alpha$ line with a rest-frame energy of $6.97$ keV corresponds to an energy-shift of $g=1.15$ or an outflow velocity of $0.2-0.5$c (depending on the half-opening angle of the funnel and thus the lamppost height, see Fig. \ref{fig:result:comparison_LP}).
Also, since we are looking directly along the jet in this system, it is likely that we are looking at the disk face on. This strongly disfavors using a standard thin disk model to explain \emph{Swift} J1644, since the Fe line can only be gravitationally redshifted from a face-on thin disk.

The observed Fe K$\alpha$ line of \emph{Swift} J1644 is illustrated together with the theoretically predicted values in the phase-space diagram in Fig. \ref{fig:result:symmetry}.
We use two different data points to illustrate \emph{Swift} J1644 in the phase-space. The two points have the same line symmetry parameter, $\tilde{w}$, but different blueshifted flux percentage, $\tilde{F}$. This is because the directly observed Fe line is very narrow, but the energy-lag analysis suggests that the line should be broader. For the broader line calculated from the lag, the blue-shifted flux percentage is $\tilde{F}=0.68$, and the width ratio is $\tilde{w}=0.85$. If we use the directly observed narrower spectral line, then it is 100$\%$ blueshifted and seems very symmetric! This might be caused by over-ionization (see Fig. \ref{fig:result:ionisation}) or  multi-scattering effects in the funnel or in the optically-thick wind. For this reason, we adopt the lower value of $\tilde{w}$ for both points. We acknowledge that the discrepancy in the line width has not been fully understood, so we link the two data points with a line to represent the uncertainty. 
Nonetheless, it can clearly be seen that both of the two \emph{Swift} J1644 data points resides in the ``super-Eddington disk zone'' on the phase-space diagram, where the  line blueshift is higher and the line shape is more symmetric compared to thin disks. A reminder is that the blueshifted flux percentage, $\tilde{F}$, is calculated assuming a $6.97$ keV rest-frame energy. Therefore, the observed blueshift of this event is even harder to be explained using thin disks.

We aim at extending the fruitful study of X-ray reverberation from thin disk models to super-Eddington accretion disk geometries.
In this letter, we have shown that the Fe K$\alpha$ fluorescent line profiles can be used to effectively distinguish super-Eddington systems from sub-Eddington systems, which is supported by the resemblance of the Fe line profiles observed from \emph{Swift} J1644 to the lines modeled here. Future studies, such as investigating the energy-lag spectrum, calculating the irradiation profile in a rigorous GR setting, experimenting with different corona and disk geometries and tracing multiple scatterings of photons in the optically thick winds, can all improve our understanding on this topic.  The study of X-ray reverberations of super-Eddington accretion flow can be widely applied to probe the funnel geometry and wind launching in TDEs, ULXs and super-Eddington Narrow-Line Seyfert 1 Galaxies, which are likely all super-Eddington systems with different Eddington ratios.
Such studies can also shed more light on how quasars in the early universe can produce winds and provide kinetic energy feedback to their host galaxies.

\acknowledgments{We acknowledge helpful discussions with Roger Blandford, Cole Miller, Sandra Raimundo, Corbin Taylor, Marianne Vestergaard, Dan Wilkins, and Yajie Yuan. We also thank the anonymous referee for constructive comments. We are indebted to the David and Lucile Packard Foundation, the Heising-Simons Foundation and the Danish National Research Foundation (DNRF132) for support. Some of the simulations carried out for this project were performed on the University of Copenhagen high-performance computing cluster funded by a grant from VILLUM FONDEN (project number 16599).}

\bibliographystyle{yahapj}
\bibliography{references}

\end{document}